# Blind Adaptive Subcarrier Combining Technique for MC-CDMA Receiver in Mobile Rayleigh Channel

Indu Shakya, Falah H. Ali, Elias Stipidis

**Abstract:** A new subcarrier combining technique is proposed for MC-CDMA receiver in mobile Rayleigh fading channel. It exploits the structure formed by repeating spreading sequences of users on different subcarriers to simultaneously suppress multiple access interference (MAI) and provide implicit channel tracking without any knowledge of the channel amplitudes or training sequences. This is achieved by adaptively weighting each subcarrier in each symbol period by employing a simple gradient descent algorithm to meet the constant modulus (CM) criterion with judicious selection of step-size. Improved BER and user capacity performance are shown with similar complexity in order of O(N) compared with conventional maximum ratio combining and equal gain combining techniques even under high channel Doppler rates.

**Introduction:** MC-CDMA is a well known multiple access technique which combines frequency-domain spreading and multicarrier modulation for effective operation in frequency selective multipath fading channel [1]. In conventional MCCDMA receivers, two subcarrier combining techniques namely, equal gain combining (EGC) and maximum ratio combining (MRC), are employed [1]. These techniques are, however, severely effected by MAI as the system loading increases, leading to poor detection performance. More advanced techniques incorporating multiuser detection, or for example, multistage RLS subcarrier-combining scheme recently proposed in [2] could be used to improve the performance. However, they are more computationally complex. A different approach is proposed in [3], that uses transmit power adaptations in time and frequency domain for MRC to reduce the degrading effects caused by weak subcarriers. However, it requires closed loop operation and very accurate channel estimation. In [4], MMSE filtering based scheme is proposed for operation in mobile Rayleigh channel for Multicarrier DS-CDMA. Although it shows much improved performance, however it requires tracking of the users' fading processes for inversion of signal correlation matrix resulting in a more complex system. In [5], a blind adaptive technique for Multicarrier DS-CDMA is proposed using a constrained constant modulus inverse QR decomposition (IQRD)-RLS algorithm for static channel conditions. The scheme is shown to outperform MRC and LMS type schemes, however, at the cost of significant increase in the complexity. In this Letter, a low complexity subcarrier combining technique is proposed for frequency-domain spread MC-CDMA that also uses the constant modulus algorithm as in [5], which is a well known algorithm in the literature. However, it is designed here in a different way that suppresses MAI with additional channel tracking capability inherent in the detection process for operation in mobile Rayleigh fading channel. More specifically, we propose an improved alternative to conventional EGC and MRC methods with similar order of complexity and without need for closed loop power adaptation strategies as in [3].

**System Model:** A synchronous model of uplink MC-CDMA similar to that in [1] is considered as shown in Figure 1. Where $b_k$ is the k[th] user's BPSK data with period of $T_b$, $c_{k,n}$ is the n[th] subcarrier chip element of the user's spreading sequence with $\sum_{n=1}^{N} c^2_{k,n} = 1$, where $N$ is the spreading length, for simplicity assumed here to be equal to the number of orthogonal subcarriers with frequencies $w_n$ radians per second. It is assumed that each subcarrier channel experiences time varying and uncorrelated (i.i.d.) Rayleigh flat fading, given by, $b_{k,n}(t) = a_{k,n}(t)e^{ij_{k,n}(t)}$, with $a_{k,n}(t)$ being the amplitude and $j_{k,n}(t)$ is the phase. The received signal $r(t)$ is given by:

$$r(t) = \sum_{m=-\infty}^{\infty} \sum_{k=1}^{K} \sum_{n=1}^{N} a_{k,n}(t) b_{k,n}(m) c_{k,n}(m) \cos\{w_n(t) + j_{k,n}(t)\} + h(t) \quad (1)$$

where $h(t)$ is the AWGN. The $r(t)$ is first subcarrier demodulated and sampled to give a signal $r_{k,n}(m)$. Coherent detection is assumed with perfect phase knowledge and real baseband model is used for the subsequent signal processing. The $r_{k,n}(m)$ is then weighted by the proposed algorithm to give an output variable $z_{k,n}(m)$ as: $z_{k,n}(m) = w_{k,n}(m) r_{k,n}(m)$, where $w_{k,n}(m)$ is a weight, normally dependent on the subcarrier combining method used to obtain the decision variable $z_k(m)$, given by $z_k(m) = \sum_{n=1}^{N} z_{k,n}(m)$. For example, in EGC, the same weight is used for all subcarriers i.e. $w_{k,n}(m) = 1 c_{k,n}$ and in MRC, a weight based on the received amplitude $a_{k,n}(m)$ is used: $w_{k,n}(m) = a_{k,n}(m) c_{k,n}$. However, in our technique the weights are determined adaptively depending upon the magnitude of previous decision variable as will be explained in next section. Finally, data estimate is obtained by deciding on $z_{k,n}(m)$, i.e. $\hat{b}_k(m) = dec\{z_k(m)\}$. For BPSK, this is simply a sign operation.

**Proposed Blind Adaptive Subcarrier Combining (BASC) Technique:** In the first symbol period $m=1$, the weights $w_{k,n}(m)$ are set as $w_{k,n}(m) = c_{k,n}(m), \forall k, \forall n$, similar to EGC. Subsequently each subcarrier signal $r_{k,n}(m)$ is weighted to obtain an output variable $z_{k,n}(m)$ as follows:

$$z_{k,n}(m) = w_{k,n}(m) r_{k,n}(m) \tag{2}$$

The decision variable $z_k(m)$ is obtained by combining the output variables as follows

$$z_k(m) = \sum_{n=1}^{N} z_{k,n}(m) \tag{3}$$

An estimate of user data is obtained as $\hat{b}_k(m) = dec\{z_k(m)\}$. The proposed algorithm seeks to maintain squared magnitude of $z_k(m)$ equal to the constant, $y, g = E\left\{\sum_{n=1}^{N} |b^k c^2{}_{k,n}| a^2{}_{k,n}\right\}$, where $E\{.\}$ is the expectation over $a_{k,n}(t)$ and for BPSK data $g = 1$. It generates error when $z^2{}_k(m) \neq g$. Since the channels vary insignificantly over a symbol period, it utilizes $z_k(m)$ and the error signal to correct weight for each subcarrier for the next symbol period, hence, attempts instantaneously to reduce the probability of generating wrong decision variable. The minimum error cost function $e_k(m)$ is evaluated as follows:

$$\min\left[ e_k(m)^2 = \{z_k(m)\{z_k(m)^2 - \boldsymbol{g}\}\}^2 \right] \qquad (4)$$

It can be observed that when $e_k(m)^2 \to 0$, our scheme approaches to the MMSE performance. Note that explicit channel tracking as in [4] is not required here. Next, error gradient signals, $\nabla_{k,n}(m)$, are calculated

$$\nabla_{k,n}(m) = e_k(m) r_{k,n}(m) \qquad (5)$$

The weights for the next symbol, $w_{k,n}(m+1)$, are updated using the gradients as follows

$$w_{k,n}(m+1) = w_{k,n}(m) - \boldsymbol{m}\nabla_{k,n}(m) \qquad (6)$$

where $\boldsymbol{m} > 0$ is the step-size. It has to be noted that, unlike in [5] where weights are used for MAI suppression only, while assuming the channels to be static, here each weight $w_{k,n}(m)$ in (6) is updated with dual purpose of suppressing MAI as well as tracking of instantaneous channel amplitude $\boldsymbol{a}_{k,n}(m)$. The accuracy of tracking and detection performance is dependent on the appropriate selection of $\boldsymbol{m}$ value and can be optimized according to the channel Doppler rate and system loading. The above processes are repeated for all $K$ users. As can be seen, the proposed algorithm has complexity in order of $O(N)$ comparable to EGC and MRC. Summary of the algorithm is also given in Table 1.

**Performance Results:** An uplink of synchronous MC-CDMA with $K$ equal power users and $N = 32$ is considered. The system is simulated in MATLAB using 10000 BPSK data symbols for each user with i.i.d. Rayleigh flat fading on each subcarrier with normalised Doppler rates of $f_d T_b = 0.003$ and $0.01$. Binary Walsh and Gold sequences are used. The latter are used here specifically to provide insight into the performance in asynchronous environments and are obtained by appending single random bits to the Gold sequences of length 31.

BER vs. $E_b/N_0$ for 10 users using Walsh and Gold sequences are shown in Figure 2 a) and b), respectively. Perfect channel estimation is assumed for the MRC in all comparisons. BASC with $m = 0.003$ is selected for reasonable adaptivity. As expected, it shows significantly improved BER compared with EGC and MRC. MRC shows degraded BER than EGC under Walsh sequences and approaches EGC in the case of Gold sequences. All schemes show less performance with Gold sequences due to higher cross-correlation. The gain in terms of number of users at $E_b/N_0 = 15 \, \text{dB}$ is presented in Figure 3 a) and b) for Walsh and Gold sequences, respectively. It can be seen that BASC gives much higher number of users. For example, at the BER of 0.01 it can support up to 19 users compared with 15 for EGC and _ 8 for MRC. Similarly BASC supports up to 17 users compared with 4 for EGC and 7 for MRC for the case of Gold sequences. In higher mobility scenario with $f_d T_b = 0.01$, BASC supports slightly less users, due to insufficient channel tracking. The effect of step-size, $m$ on the BER is also investigated in Figure 4 for $K = 10$ and 16 under the same conditions: $E_b/N_0 = 20$ dB, $f_d T_b = 0.003$, and $m = 0.003 - 0.04$. It can be seen that BASC provides better performance for all step-size considered. It is found that the optimum step-size are $m = 0.025$ and 0.02.

**Conclusion:** A low complexity subcarrier combining technique for MC-CDMA receiver is presented in mobile Rayleigh channel. It is shown that, by exploiting the structure formed by users' repeating sequences into an adaptive algorithm with simple CM cost function and judicious selection of step-size, simultaneous supersession of MAI and implicit channel tracking is achieved without any knowledge of the channel. Improved BER and user capacity compared with MRC and EGC are shown. For example, at the BER of 0.01, it can support 17 users compared with 7 and 4 users for MRC and EGC, respectively. Optimum step-size are also investigated using simulations to give more improved performance.

**Submitted to IET Electronics Letters**

**Authors' Affiliations:** Indu Shakya, Falah H. Ali, Elias Stipidis (Communications Research Group, Department of Engineering and Design, University of Sussex, Brighton, BN1 9QT, UK, Email: f.h.ali@sussex.ac.uk)

**Figure Captions:**

Figure 1: Proposed MC-CDMA system model

Figure 2: BER vs. $E_b/N_0$ performance of the proposed BASC for K=10 and $\mu = 0.003$, for a) Walsh sequences and b) Gold sequences of N=32

Figure 3: BER vs. number of users performance of the proposed BASC for Eb/N0 = 15 dB and μ = 0.003, for a) Walsh sequences and b) Gold sequences of N=32

Figure 4: The effect of step-size, μ on the BER performance of the proposed BASC for K = 10 and 16 under Eb/N0 = 20 dB, fdTb = 0.003 and Walsh sequences of N=32

Figure 1:

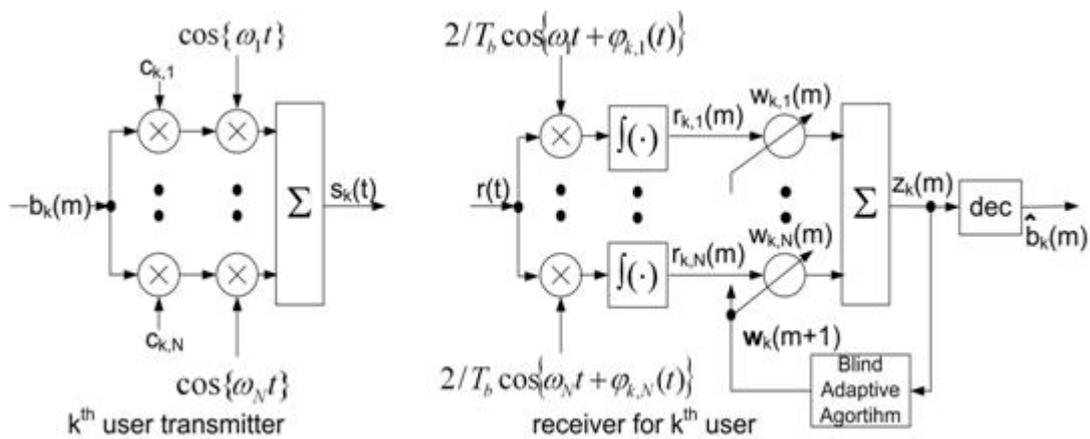

| Table 1. The proposed BASC algorithm steps |
|---|
| Set the values: $m$, and $g$ |
| At $m=1$, initialize $w_{k,n}(m) = c_{k,n}(m), \forall k, \forall n$ |
| For $m = 1,2,3,...$, carry out steps 1 - 7 |
| 1 Perform weighting, $z_{k,n}(m) = w_{k,n}(m) r_{k,n}(m)$ |
| 2 Combine signals, $z_k(m) = \sum_{n=1}^{N} z_{k,n}(m)$ |
| 3 Obtain data estimate, $\hat{b}_k(m) = dec\{z_k(m)\}$ |
| 4 Evaluate, $\min\left[ e_k(m)^2 = \left\{ z_k(m)\{z_k(m)^2 - g\} \right\}^2 \right]$ |
| 5 Calculate gradients, $\nabla_{k,n}(m) = e_k(m) r_{k,n}(m)$ |
| 6 Update weights, $w_{k,n}(m+1) = w_{k,n}(m) - m\nabla_{k,n}(m)$ |
| 7 Stop after the detection of all K user |

Figure 2:

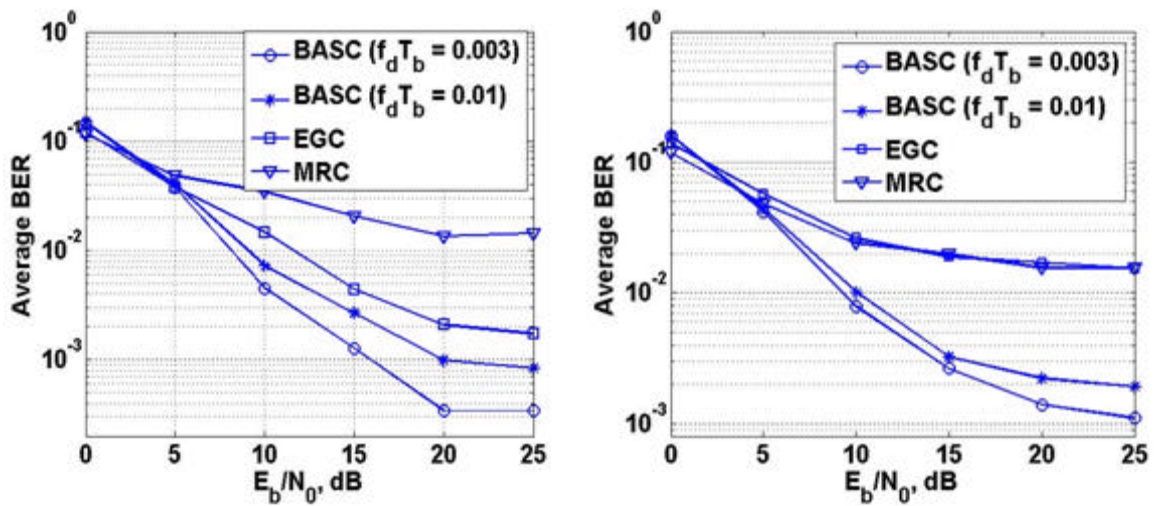

Figure 3:

**Submitted to IET Electronics Letters**

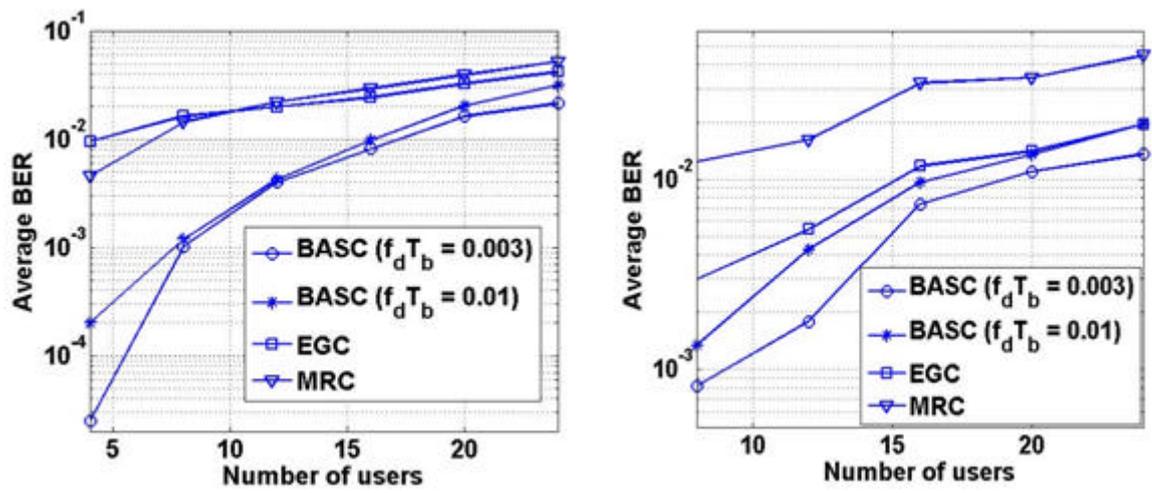

Figure 4:

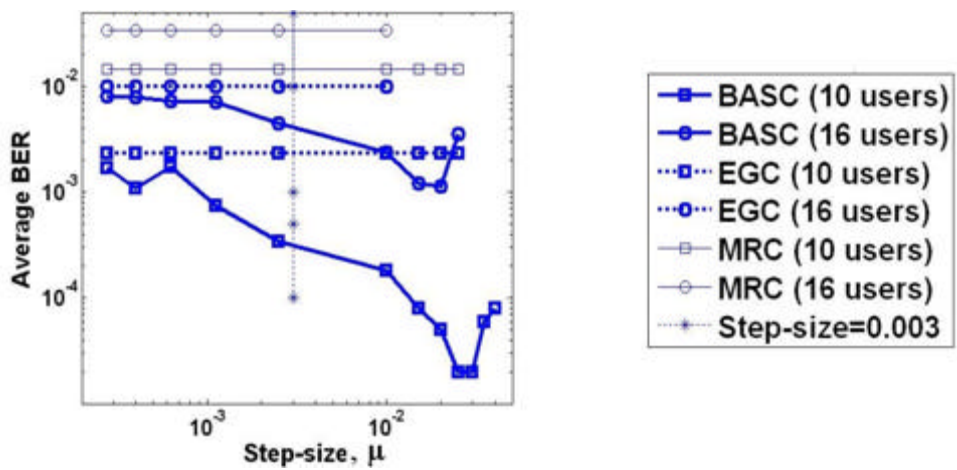